\documentclass[twocolumn,showpacs,preprintnumbers,amsmath,amssymb]{revtex4}
\usepackage{graphicx}
\usepackage{amsmath}
\usepackage{times}
\usepackage{color}

\begin{document}
\title{Congestion diffusion and decongestion strategy in networked traffic}
\author{Zhi-Xi Wu}
\affiliation{Department of Electronic Engineering, City University
of Hong Kong, Kowloon, Hong Kong, China}
\author{Wen-Xu Wang} \affiliation{Department of Electronic Engineering, City University
of Hong Kong, Kowloon, Hong Kong, China}
\author{Kai-Hau Yeung} \email{eeayeung@cityu.edu.hk}
\affiliation{Department of Electronic Engineering, City University
of Hong Kong, Kowloon, Hong Kong, China}
\date{Received: date / Revised version: date}
\begin{abstract}
We study the information traffic in Barab\'asi-Albert scale free
networks wherein each node has finite queue length to store the
packets. It is found that in the case of shortest path routing
strategy the networks undergo a first order phase transition i.e.,
from a free flow state to full congestion sate, with the
increasing of the packet generation rate. We also incorporate
random effect (namely random selection of a neighbor to deliver
packets) as well as a control method (namely the packet-dropping
strategy of the congested nodes after some delay time $T$) into
the routing protocol to test the traffic capacity of the
heterogeneous networks. It is shown that there exists optimal
value of $T$ for the networks to achieve the best handling
ability, and the presence of appropriate random effect also
attributes to the performance of the networks.

\end{abstract}
 \maketitle

\section{Introduction}
In the last few years, complex networked systems have attracted
much attention, ranging from the areas of physics, sociology,
biology, to technology and many others
\cite{Albert2002rmp,Newman2003siam,Boccaletti2006pr,Pastor2004book,
Newman2006book}. It has been well proved that the topological
features of underling interaction networks have great impacts on
the final outcomes of the dynamics taking place on them
\cite{Albert2002rmp,Newman2003siam,Boccaletti2006pr,Pastor2004book,
Newman2006book}. For example, the scale free topology of a network
results in a vanishing threshold of epidemic spreading on it with
the increase of the network size \cite{Pastor2001prl}, also gives
rise to a robust behavior against random failures and fragile for
aimed attack \cite{Albert2000nature}; the networks with
homogeneous degree distribution, small average path length, and
small clustering coefficient are found to be more easily to retain
synchronization \cite{Nishikawa2003prl}.

Among all spreading and transport problems (epidemic, opinion,
cascading, information, etc.) investigated in various kinds of
complex networks
\cite{Newman2003siam,Boccaletti2006pr,Newman2006book,Ohira1998pre,
Germano2006pre,Galam1999pa,Sznajd2000ijpc}, the information
transport in communication networks such as the Internet may be of
practical importance \cite{Barthelemy}, since it plays a more and
more important role in our daily life, e.g., data resources, the
e-business, online games, and so may others. The ever-increasing
number of users of the Internet, hence the need of tremendous
amount of information transport make it necessary study how the
topological properties of the underlying infrastructures influence
the traffic flow \cite{Tadic2002}. In the studying of information
traffic in complex networks, the following two problems are mostly
focused on
\cite{Guimera2002prl,Cholvi2005pre,Danila2006pre,Danila2007chaos,
Sreenivasan2007pre,Yan2006pre,wang2006pre,Liu2007pre}: Given the
underlying network, what is the efficient routing protocol to
optimize the packet delivery; and given a routing protocol, what
type structure of the network is optimal, i.e., the performance of
the structure or evaluation of the effect of structure on the
traffic (optimality is defined as the minimization of the average
packet arrival time and the maximization of the packet handling
capacity of the network). In this paper, we would like to
investigate in detail how the underling topological structure
affects the data traffic in complex heterogeneous networks.

All too often, the buffer size of the nodes are assumed to be
infinite, i.e., all nodes can receive data packets as many as
possible
\cite{Guimera2002prl,Cholvi2005pre,Danila2006pre,Danila2007chaos,
Sreenivasan2007pre,Yan2006pre,wang2006pre,Liu2007pre}. However,
due to the physical constraint, we know that all data-processing
machines have a finite buffer (or queue length, denoted by $L$ in
the remaining parts for simplicity) to store data packets. It is
reasonable and necessary to take into account this fact in the
traffic study. So, in the present paper, we discard the infinite
buffer assumption, but rather consider finite buffer of each node,
and study how this restraint affect the data traffic. All of our
results in the following parts are presented for $L=5$. Other
selection of value of $L$ does not change the qualitative behavior
of the results shown below. In addition we also incorporate random
effect as well as a control method into the routing protocol to
study the performance of the networks. It is shown that due to the
finite storage of the nodes, a first order phase transition
emerges in the information transportation process. The random
effect and the control method in the routing protocol are also
found to have a great influence on the handling capacity of the
networks.


\section{The model}

It has been proposed that scale free network topology is a
suitable candidate for the structure of the internet at AS level
\cite{Capocci2001pre,note}. For simplicity, we use the well-known
Barab\'asi-Albert (BA) scale-free network model
\cite{Barabasi1999science} as the physical infrastructure on top
of which a packet delivery process is taking place. The BA model
contains two generic mechanisms of many real complex systems:
growth and preferential attachment
\cite{Barabasi1999science,Albert2002rmp}, which can be constructed
as follows. Starting from $m_0$ nodes, one node with $m$ links is
attached at each time step in such a way that the probability
$\prod_i$ of being connected to the existing node $i$ is
proportional to the degree $k_i$ of that node, i.e., $\prod_i=k_i
/\sum_jk_j$, where $j$ runs over all existing nodes. In the
present work, the total network size is fixed as $N=1000$ and the
parameters are set to be $m_0=m=3$ (hence the average connectivity
of the network is $\langle k\rangle=6$ \cite{Albert2002rmp}). The
degree distribution of the generated BA network $P(k)$, which
denotes the probability of a randomly selected node in the network
having exactly degree $k$, follows a power law $P(k)\sim
k^{-\gamma}$ with the exponent $\gamma=3$ in the large degree
limit.

Each node of the underlying infrastructure acts both roles of a
host and a router at the same time. Due to the finite storage
capacity of the nodes, we now define that a node is congested if
its buffer is fully filled by packets. The packet transmission on
the network, i.e., the packet generating process and the packet
delivery process, is implemented by a discrete time parallel
update algorithm. At each time step, the probability for node $i$
to generate a packet is $R$ if there are some free place in its
buffer (i.e., available of its queue), otherwise no new packet is
inserted. Once a packet is created, its destination is chosen
uniformly at random among the other $N-1$ nodes of the network.
Each newly inserted packet is placed at the end of the queue of
the node which generates it. After all the nodes have finished the
packet generating process, they start to deliver the packets
stored in their queue, which are composed of all packets that were
sent to them by their neighbors in the previous steps and packets
created by themselves (if any). For simplicity, and without loss
of generality, we assume that all nodes have the same processing
capacity of one data packet per time step. The first packet in the
queue is then sent to a neighbor following some routing protocol.

Although in modelling communication networks like the Internet,
the routing process following the shortest path (SP) from a given
source to its destination is usually preferable, many previous
studies have shown that some certain degree of stochasticity can
to some extent enhance the traffic handling capacity of the
underlying network
\cite{Echenique2004pre,Zhao2005pre,Ashton2005prl,Jarrett2006pre}.
On the other hand, it has been found that for the internet almost
$90\%$ data packets are forwarded along their SP from source to
destination, and the deviation of traffic from the SP is only
about $10\%$ \cite{Krioukov2003arxiv}. Inspired by these two
factors, we introduce in our model some randomness for packet
delivery process, namely, we let data packets be delivered along
their SP from source to destination with a probability $(1-p)$
\cite{note1}, and be sent to a randomly neighbor with probability
$p$. Whenever a packet is forwarded along SP, if there exist
several shortest paths, we just select randomly one of them to
deliver the packet. For $p=0.0$, we recover the SP delivery
protocol. For other values of $p$ greater than zero, we
incorporate random effect into the packet delivery. To make a
close correlation to the realistic scenario of data traffic, here
we restrain our studies for the values of $p$ between $0.0$ and
$0.2$. Once a packet reaches its destination through the above
routing protocol, it will be removed from the system.

We want to remark that the packet generation rate $R$ in our model
is different from that in the case of the infinite queue length
\cite{Guimera2002prl,Cholvi2005pre,Danila2006pre,Danila2007chaos,
Sreenivasan2007pre,Yan2006pre,wang2006pre,Liu2007pre}. In fact,
since those congested nodes (whose buffer are fully filled)
temporarily can not generate any new data packets, the
\emph{effective} average packet generation rate of the whole
network is equal or less then $R$. Nevertheless, we also denote
average packet generation rate by $R$ for convenience.

\section{Results and discussion}

\begin{figure}[h]
\begin{center}
{\includegraphics[width=8cm]{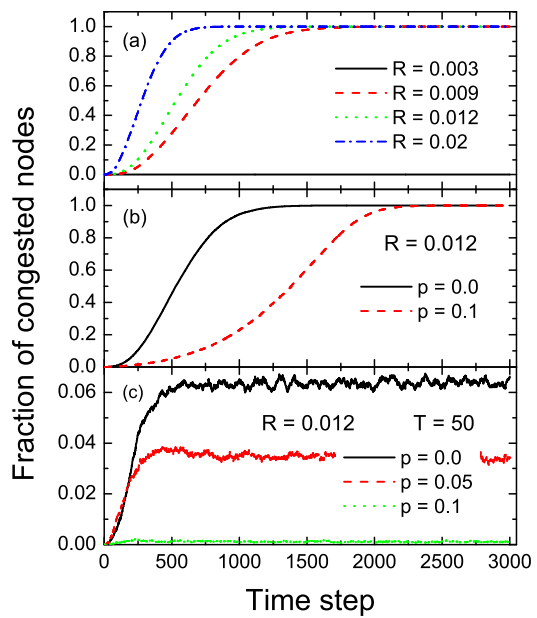}} \caption{Time evolution of
the congested nodes in the BA scale free networks of size $N=1000$
with $m=m_0=3$. (a) is for the data packets forwarded by using of
SP routing protocol. (b) is for the case of some random effect
involved in the packet delivery process. (c) is for the case of
packet dropping strategy applied to the congested nodes. (see the
text for details)} \label{fig1}
\end{center}
\end{figure}

According to the above algorithm for packet delivery, one can
expect that for large values of $R$, the whole network would fall
easily into a fully congested state due to the finite queue length
of the nodes. In Fig. \ref{fig1}(a), we show four typical time
evolution series of the average fraction of congested nodes in the
BA scale free networks with total size $N=1000$ and average
connectivity $6$. The different styled lines correspond to
different packet generation rate $R=0.003$ ,$0.009$, $0.012$ and
$0.02$, respectively. From Fig. \ref{fig1}(a) one can see that for
sufficient small $R$, e.g., $R=0.003$, the networks sustain a free
flow state (absence of congestion), while for large values of $R$,
the networks are doomed to jam and the fraction of congested nodes
in the networks increase very fast with the increment of time
steps. The greater the value of $R$ is, the more sharply the
curves ascend. We have also implemented simulations for other
values of $N$, $\langle k\rangle$ and $L$. The qualitative
behaviors of the time evolution curves shown in Fig. \ref{fig1} do
not change (not shown here).

In Fig. \ref{fig1}(b), the time evolution of the average fraction
of congested nodes is plotted for a special value of $R=0.012$
with the incorporation of random effect in the routing protocols
$(p=0.1)$. For comparison, the result for $p=0.0$ is also shown.
It is clear the curve for $p=0.1$ ascends more slowly than that
for $p=0.0$, which indicates that some appropriate degree of
stochasticity for routing protocols can improve the handling
ability of the network in accordance with previous observations in
the case of infinite queue length
\cite{Echenique2004pre,Zhao2005pre,Ashton2005prl,Jarrett2006pre}.
Despite of this point, we can see the network is still doomed to
jam in the long time limit, which is also caused by the finite
buffer size of the nodes. (Fig. \ref{fig1}(c) will be discussed
later)

\begin{figure}[h]
\begin{center}
{\includegraphics[width=8cm]{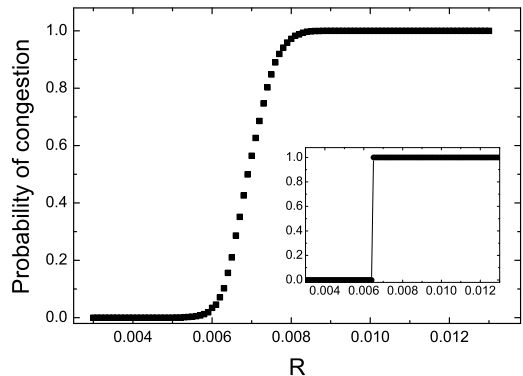}} \caption{Probability of
congestion \emph{Vs.} the data packet generation rate $R$ in the
BA networks of size $N=1000$ with $m=m_0=3$. Each data point is
obtained by averaging over 15000 samples (a set of 50 realizations
of the BA network, and 300 independent experiments for each of
them). The inset shows the result of a typical experiment of
information traffic in a BA network. The Y-axis of the inset
denotes the fraction of congested nodes.} \label{fig2}
\end{center}
\end{figure}

Since whether the underlying networks are congested or not depends
strongly on the packet generation rate $R$, we would first like to
investigate the behavior of the congestion as a function of $R$.
In the inset of Fig. \ref{fig2}, we show the result of a typical
sample of the information traffic in a BA network. The packets are
forwarded by using the SP routing protocol. It is obvious that a
first order phase transition emerges at a certain value of $R_c$
above which the network is doomed to congestion. Due to the finite
size of the considered BA network ($N=1000$ in the present case),
the value of $R_c$ for the traffic in different network
realizations will vary to some extent. To obtain a more precise
picture of the phase diagram, we have implemented $15000$ samples
including a set of $50$ realizations of the BA network, and $300$
independent experiments on each of them. We calculated the total
number of times that the networks fall into full congestion by
varying the value of $R$, and then the results are normalized by
the total number of samples. The final results, ie., the average
probability of congestion as a function of $R$, are plotted in the
main panel of Fig. \ref{fig2}, and the threshold $R_c$ is about
$0.005(5)$ in the present case (obviously, for other values of
$N$, $m_0$, $L$, this value will vary).

\begin{figure}[h]
\begin{center}
{\includegraphics[width=8cm]{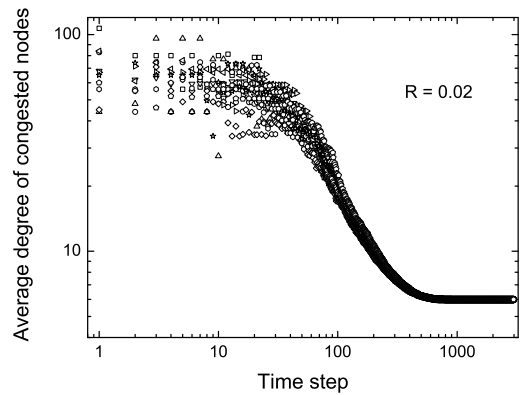}} \caption{Time behavior of
the average degree of the congested nodes for congestion outbreaks
($R=0.02>R_c$) in BA networks of size $N=1000$ with $m=m_0=3$. The
shown results are for ten independent samples (plotted by
different symbols).} \label{fig3}
\end{center}
\end{figure}

From Fig. \ref{fig1} (a), we can see that, once some nodes are
jammed, the congestion spreads very fast in the whole network if
no control measure is deployed. From point view of developing
control strategies to prevent the whole network from malfunction,
it is valuable to figure out the detail knowledge of the way that
the congestion spreads through the network. Since the packets are
delivered according to the SP protocol, the hubs (having large
degrees and hence high betweenness) are more prone to congestion
than those nodes with small degrees. A simple way of
characterization of the congestion diffusion through the network
is to study a convenient quantity, namely the average degree of
congested nodes (if any), in numerical transporting experiments.
We show in Fig. \ref{fig3} this quantity for ten independent
samples in the BA networks as a function of simulation time step.
It is clear that the curves show an initial plateau whose values
are far greater than the average connectivity of the network $6$,
which indicates undoubtedly that the large degree nodes in the
network are preferentially congested. After the hubs are blocked,
those nodes with largest degree in the remaining are successive to
be congested, which is reflected by the smooth decreasing of the
curves. Taken together, the dynamical spreading process of
congestion is clear: with more and more packets inserted in the
network, the hubs with high betweenness are firstly congested, and
then the congestion spreads going towards those nodes with smaller
and smaller degrees. Finally, the whole network is totally jammed,
and the average degree of congested nodes is no other than the
average connectivity of the BA networks.

From the above scenario of the congestion propagation in
heterogeneous networks, one may realize that an efficient way to
prevent the whole network from malfunction is to detect the
blocked nodes in the very early stage of congestion, and to deal
with the problem as soon as possible. However, due to the
fluctuation of information flow in real communication systems
\cite{deMenezes} as well as many other physical restraints, it is
inconvenient and even impossible to make immediate measurements
and treatments on the congested nodes before a large scale
breakdown of the system. More easier is to let the congested nodes
themselves have the ability of getting out of jam in terms of some
appropriate instructions. As an alternative way, in the following
part we implement a packet-dropping strategy for the congested
nodes to prevent the whole network from jamming.


\begin{figure}[h]
\begin{center}
{\includegraphics[width=8cm]{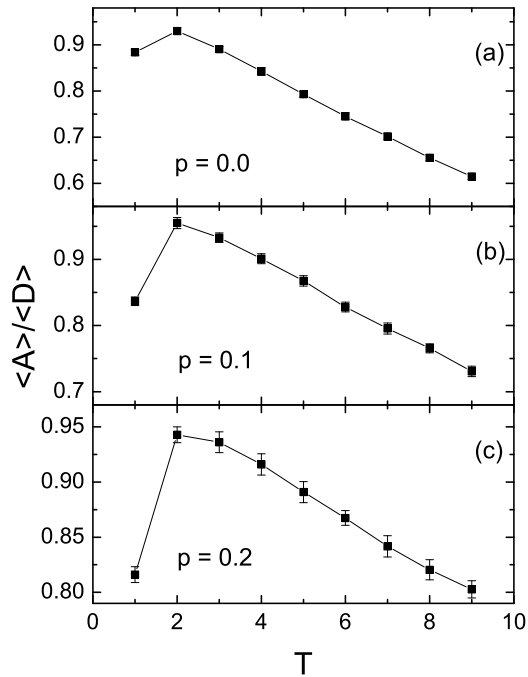}} \caption{The ratio of the
average number of successfully arrived packets, $\langle
A\rangle$, to the average number of dropped packets, $\langle D
\rangle$, as a function of the delay time $T$ for different values
of $p$. The error bar denotes the standard deviation from a total
of 30 traffic simulations on one realization of BA network. The
parameter $R$ is equals to 0.05.} \label{fig4}
\end{center}
\end{figure}

We assume that if a node is congested and the persistence of this
congested state exceeds some time $T$, then the node will simply
empty its buffer to regain the capacity of receiving and handling
packets. In practice, this method can be easily implemented by
using control software, for example, the long time congested
router can call for a pre-installed instruction to empty its
buffer. In our computer simulations, if the congested state of a
node lasts for more than $T$ time step (for convenience, the time
that a node is just congested is marked by $T=1$), we reset its
buffer size back to $L=5$, and those stored packets in its buffer
are discarded. Obviously, by doing so, we face the problem that
many packets will be lost and disappeared forever in the network.
On the other hand, however, if no strategy such as packet-dropping
is implemented, the whole network would fall rapidly into
collapse. On balance, an acceptable method is to select an
appropriate parameter $T$ to achieve as least packet-losing as
possible, while keeping the whole network functional.

In Fig. \ref{fig1}(c), we show the time evolution of the fraction
of congested nodes for $T=50$ and several values of $p$. The
packet generation rate $R$ is set to $0.012$, which guarantees
that the whole network is blocked without the implement of the
strategy of packet-dropping [as displayed in Fig. \ref{fig1}(b)].
From Fig. \ref{fig1}(c), we note that even for so large value of
$T$, the networks remain at low level congestion, and if some
random effects are involved in the packet delivery, the congestion
level may further decrease to a negligible level [the curve for
$p=0.1$ in Fig. \ref{fig1}(c)]. If $T$ is too large, we know that
the whole network is still easily to be jammed for large values of
$R$; if $T$ is too small, however, there would be so many packets
losing. Take both into consideration, there should be an
intermediate value of $T$ to achieve optimal performance. To
evaluate the efficiency of the packet dropping method, we study
the ratio of the average number of successfully arrived packets
$\langle A\rangle$ to the average number of dropped packets
$\langle D\rangle$ by tuning the values of the parameters $T$, $p$
and $R$.

\begin{figure}[h]
\begin{center}
{\includegraphics[width=8cm]{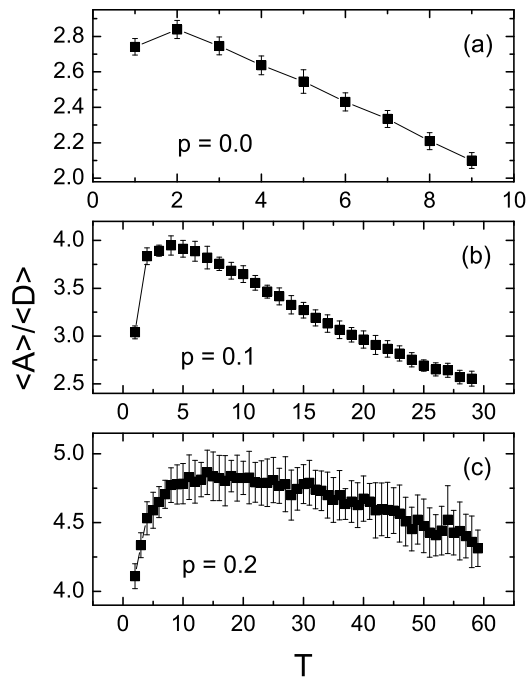}} \caption{As shown in Fig.
\ref{fig4}, but for $R=0.02$.} \label{fig5}
\end{center}
\end{figure}

\begin{figure}[h]
\begin{center}
{\includegraphics[width=8cm]{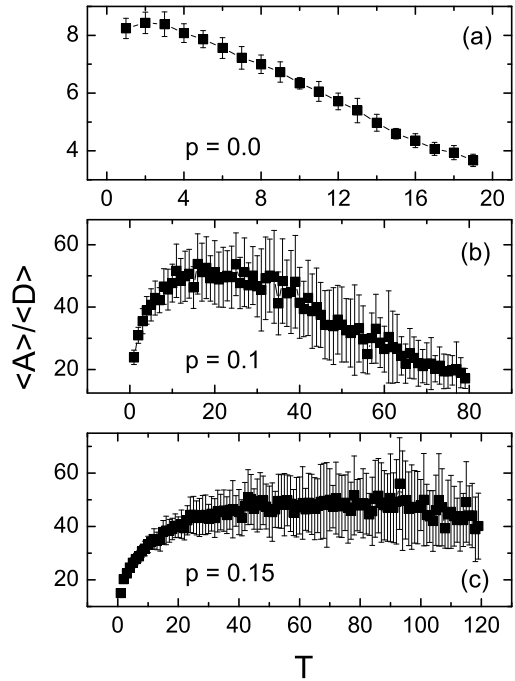}} \caption{As shown in Fig.
\ref{fig4}, but for $R=0.012$.} \label{fig6}
\end{center}
\end{figure}

In Figs. \ref{fig4}-\ref{fig6} we show the quantity $\langle
A\rangle/\langle D\rangle$, respectively, as a function of $T$ for
several values of $p$ and $R$. The presented simulation results
are obtained by averaging over $30$ independent traffic
realizations on the BA scale free networks of total size $N=1000$
and $m=m_0=3$. For large packet generating rate $R=0.05$ (Fig.
\ref{fig4}), there exists an optimal value for $\langle
A\rangle/\langle D\rangle$ at $T=2$, which is insensitive to the
detailed value of $p$. This result indicates that at times of high
flux (large values of $R$), the most efficient way to alleviate
traffic congestion and sustain the overall traffic handling
ability of the heterogeneous networks is to empty the buffers of
those congested nodes very immediately. However, it is worthy
pointing out that the simulation results in Fig. \ref{fig4} show
clearly that the most appropriate time for the congested nodes to
empty their buffer is not the time that they are just jammed, but
one more time step after they are congested. This perhaps is due
to the high rate of packets loss induced by immediate dropping of
the packets stored in the congested nodes' buffer, which otherwise
may be delivered in the next time step to their unblocked
neighboring nodes.

The above picture, however, is changed for smaller (yet greater
than the $R_c\approx 0.005$) values of $R$ (Figs. \ref{fig5} and
\ref{fig6}). For $p=0$, i.e., the packets are forwarded by using
of the SP protocol, the most appropriate time for the congested
nodes to empty their buffer is the same as before: at $T=2$,
namely one more time step after they are congested [Figs.
\ref{fig5}(a) and \ref{fig6}(a)]. When there exist some stochastic
effects in the packet delivery, i.e., for any value of $p$ greater
than $0$, the most appropriate time for the congested nodes to
empty their buffer shows nontrivial behavior, which depends
closely both on the values of $R$ and $p$. More precisely, the
smaller the value of $R$ as well as the larger the value of $p$,
the larger the time $T$ is suitable [Figs. \ref{fig5} (b) and (c),
Figs. \ref{fig6} (b) and (c)]. This means that at times of low
flux, in order to achieve high packet arrival rate, the right way
is just to let those congested node remain their blocked state for
some appropriate time. It is worthy stressing that the condition
of $p>0$ should be satisfied, which is the actual case in
realistic communication systems \cite{Krioukov2003arxiv}. Finally,
we want to remark that the curves shown in Figs. \ref{fig5} and
\ref{fig6} are obtained from averages of 30 independent traffic
experiments on one BA network realization. However, we have
checked that for another independent realization of the underlying
infrastructure, the shape of the curves may deviate to some
extent, but all the qualitative properties of them (just as was
shown in Figs. \ref{fig5} and \ref{fig6}) remain absolutely
unchanged.


\section{Conclusion}
In summary, we have studied the information traffic in BA scale
free heterogeneous networks. The nodes are endowed with finite
buffer size and the same capacity of processing packets in each
time step. It was found that for sufficiently small packet
generation rate, the networks sustain a free flow state, while for
large packet generation rate, the underlying infrastructures
sooner or later fall into totally jammed. The phase transition for
the heterogeneous networks going through the free flow state to
the fully congested state with the increment of packet generation
rate is of first order. Whenever the heterogeneous networks are
going to be jammed, we have shown that the congestion takes first
control of the hub nodes in the networks, and then it rapidly
invades the whole network via a hierarchical cascade through
progressively the nodes with smaller and smaller degrees. We have
also investigated a control method, namely the packet-dropping
strategy with a delay time $T$, to keep the whole network
functional in the case of high packet generation rate. It was
found that by using of this simple strategy, a network can sustain
well its packet handling ability at the expense of lowering packet
arrival rate. The efficiency and quality of the control method is
determined by some correlated factors, e.g., the magnitude of the
packet generation rate and the degree of stochasticity of the
routing protocols as well.


\acknowledgments{We thank our colleague, Sammy Chan, for useful
discussions and careful reading of the manuscript. This work was
supported by City University Strategic Research Grant numbered
7001941.}

\bibliographystyle{h-physrev3}

\begin{thebibliography}{10}
\bibitem{note1}
Here, we simly define the SPs of any given two nodes as the paths
with minmal number of hops when going from one to the other,
wheareas it is not always the actual case in realistic situation.
\end{thebibliography}

\end{document}